\documentclass[aps,twocolumn,showpacs, amsmath,amssymb,prstab]{revtex4}
\usepackage{graphicx}
\usepackage{bm}

\begin{document}

\newcommand{\D}{\,\mathrm{d}}
\newcommand{\I}{\mathrm{i}}

\title{Synchrotron dynamics in Compton x-ray ring with nonlinear momentum compaction}

\author{Eugene Bulyak}
\email{bulyak@kipt.kharkov.ua}
\author{Peter Gladkikh}%
\author{Vladislav Skomorokhov}%
\affiliation{%
NSC KIPT  
Kharkov, Ukraine
}%

\date{\today}

\begin{abstract}
The longitudinal dynamics of electron bunches with a large energy
spread circulating in the storage rings with a small momentum
compaction factor is considered. Also the structure of the
longitudinal phase space is considered as well as its modification
due to changes in the ring parameters. The response of an
equilibrium area upon changes of the nonlinear momentum compaction
factor is presented.
\end{abstract}


\maketitle

\section{Introduction}
Engagement of electron storage rings for production of x rays
through Compton scattering of laser photons against
ultrarelativistic electrons was proposed in 1998 \cite{huang98}. Two
basic schemes exist so far. One of them supposes use of electron
beams with unsteady parameters \cite{loewen} and applies the
continual injection (and ejection of circulating bunches by the next
injecting pulse) of dense intensive bunches. The second scheme is
based on the continuous circulation of bunches. To make the bunches
acquired a sufficiently large energy spread confined (see
\cite{buepac04}), a lattice with a small controllable momentum
compaction factor is proposed to employ \cite{gladkikh05}. The
longitudinal dynamics in the small compaction lattice is governed
not only by the linear effects of the momentum deviation but by the
nonlinear ones as well.

In Compton sources storing the bunches with the large energy spread
which can be as high as a few percents, ring's energy acceptance
becomes compared to the energy spread. To get proper life time of
the circulating electrons, the energy acceptance $\sigma\equiv \max
(E-E_s)/E_s$ ($E_s$ is the energy of synchronous particle) should be
high enough.

Within a linear approximation according to the energy deviation, the
acceptance can be increased either by enhancement of the RF voltage,
$V_\mathrm{rf}$, or by decreasing of the linear momentum compaction
factor $\alpha_0$ since
$\sigma\propto\sqrt{V_\mathrm{rf}/\alpha_0}$.

The paper presents results of study on the longitudinal dynamics of
electron bunches circulating in storage rings with a small linear
momentum compaction factor $\alpha_0$. Structure of the phase space
are considered and its deformation with changes in the ring lattice
parameters. In particular, the size of stable area as a function of
the RF voltage and momentum compaction is evaluated.

\section{Finite-difference model of longitudinal motion}
Let us consider a model of the ring comprised only two components:
drift and radio frequency (rf) cavity. For the sake of simplicity we
will suggest the cavity infinitely short, in which the particle
momentum (energy) suffer an abrupt change while the phase of a
particle remains unchanged. On the contrary, the phase of a particle
traveling along the drift changes while the energy remains
invariable. The longitudinal motion in a such idealized ring will be
described in canonically conjugated variables $\phi $ (the phase
about zero voltage in the cavity) and the momentum $p\equiv (\gamma
-\gamma_s)/\gamma_s$ equal to the relative deviation of the particle
energy from the synchronous one ($\gamma_s $ is the Lorentz factor
of the synchronous particle).

To study systems able to confine the beams with large energy spread,
one needs to account not only the linear part of the orbit deviation
from the synchronous one, but nonlinear terms as well:
\begin{equation} \label{c:1a}
\Delta x\approx D_1p+D_2p^2+\dots \;,
\end{equation}
where $D_1$ and $D_2$ are the dispersion functions of the first and
second orders, respectively.

Accordingly, relative lengthening of a (flat) orbit is
\begin{align} \label{c:2}
\frac{\Delta L}{L_0}&=\oint\sqrt{\left(1+\frac{\Delta x}{\rho}
\right)^2+\left(\frac{\D \Delta x}{\D s} \right)^2}\D s\nonumber\\
& \approx \alpha_0p +\alpha_1p^2+\dots\; ,
\end{align}
where $L_0$ is the length of synchronous orbit, $\rho(s)$ the local
radius of curvature, $s$ the longitudinal coordinate. The
coefficients $\alpha_0$ and $\alpha_1$ are determined as
\begin{subequations}\label{c:3}
\begin{align}
\alpha_1 &= \frac1{L_0}\oint\frac{D_1}{\rho}\D s\; ;
\label{c:3a}\\
\alpha_2 &= \frac1{L_0}\oint\left(\frac{D'^2_1}{2}+
\frac{D_2}{\rho}\right)\D s \;.
\end{align}
\end{subequations}

In accordance with the definitions for $\alpha_0$ and $\alpha_1$,
the momentum compaction factor $\alpha_c$ can be written as
\begin{equation} \label{c:1}
\alpha_c = \frac1{L_0}\frac{\D L}{\D\delta}\approx \alpha_0
+2\alpha_1 p + \dots \;.
\end{equation}

To study the phase dynamics in a storage ring with small momentum
compaction factor $\alpha_0$, the next terms of expansion of the
compaction over the energy deviation should be accounted for, hence
--- the higher terms in the sliding factor $\eta$
\cite{pellegrini91,lin93,feikes04}. Magnitude of $\eta$
characterizes a relative variation of the phase due to changes of
the particle velocity and orbit length. It is determined by the
relation
\begin{equation} \label{c:4}
\frac{\Delta\phi}{\phi}= \eta(p) p \approx (\eta_0 + \eta_1 p +
\dots)p\;,
\end{equation}
with $\eta_0$ and $\eta_1$ having been determined by
\begin{subequations}\label{c:4a}
\begin{align}
\eta_0 &=\alpha_0-1/\gamma^2_s\\
\eta_1 &=\alpha_1+\eta_0+\frac{3}{2\gamma^2_s} \left(
1-\frac{1}{\gamma^2_s} \right) \; .
\end{align}
\end{subequations}

The finite-difference equations for the phase $\phi$ and the
variation of relative energy $p$ in the model under consideration
read
\begin{subequations} \label{eq:3}
\begin{align}
\phi_f&= \phi_i+ (\kappa_0 p_i+\kappa_1 p_i^2)\Delta\tau\; ;\label{eq:3a} \\
p_f&= p_i -U_\mathrm{rf}\sin\phi_f \Delta\tau\; \label{eq:3b},
\end{align}
\end{subequations}
%
where
\begin{equation} \label{c:5}
\Delta\tau=\tau_f-\tau_i=\frac{\beta c}{L}(t_f-t_i)\nonumber\;,
\end{equation}
the subscripts $i$ and $f$ correspond to the initial and final
values, respectively. The dimensionless variable $\tau=t\beta c/L$
represents time expressed in number of rotations ($t$ is time,
$\beta c$ the velocity of a particle). The factors $\kappa_0$ and
$\kappa_1$ at a large $\gamma_s$ are determined by the expressions
$\kappa_0 = 2\pi h \eta_0\approx 2\pi h \alpha_0$, $\kappa_1 = 2\pi
h \eta_1\approx 2\pi h (\alpha_0+\alpha_1)$ ( $h$ the harmonic
number).

From Eqs.~\eqref{eq:3}, differential (smoothed) equations can be
deduced. As it seen, the RHS of \eqref{eq:3b} contains the final
value of the phase $\phi_f$ expressed via the initial value $\phi_i$
and momentum $p_i$ by the equation \eqref{eq:3a}.

Let us expand $\sin\phi_f$ into series of powers of $\Delta\tau$:
\begin{align} \label{c:6}
\sin\phi_f &= \sin\left[ \phi_i+ (\kappa_0 p_i+\kappa_1
p_i^2)\Delta\tau\right]\nonumber\;\\
&\approx \sin\phi_i+\cos\phi_i(\kappa_0 p_i+\kappa_1
p_i^2)\Delta\tau\; .
\end{align}

Since $\Delta\tau$ can not be regarded as infinitesimal (formally
Eqs.~\eqref{eq:3} present a complete turn, $\Delta\tau = 1$), then
the linear term can be neglected if $\kappa_0 p_i + \kappa_1
p_i^2\ll 1$. In the considered case it can be done since maximum of
the energy spread does not exceed a few percents, and the momentum
compaction factor $\alpha_0$ supposed small. From these assumptions,
finite difference equations reduce to
\begin{subequations} \label{eq:3c}
\begin{align}
\frac{\Delta\phi}{\Delta\tau}&=\kappa_0 p_i+\kappa_1 p_i^2\; ; \\
\frac{\Delta p}{\Delta\tau}&= -U_\mathrm{rf}\sin\phi_i \; .
\end{align}
\end{subequations}

\section{Differential model of motion}
Noting of formal similarity of Eqs.~\eqref{c:6} to canonical
Hamilton equations describing a mathematical pendulum, we can use a
smoothed analog to these equations (a differential substitute for a
finite difference equation, $\Delta\tau \to 0$) to facilitate
analysis of the motion
\begin{subequations}\label{eq:4}
\begin{align}
\frac{\D\phi}{\D\tau} &= \kappa_0 p+\kappa_1 p^2\; ; \\
\frac{\D p}{\D\tau} &= -U_\mathrm{rf}\sin\phi\; .
\end{align}
\end{subequations}

A Hamilton function for \eqref{eq:4} possesses a specific form with
the cubic canonical momentum term
\begin{equation} \label{eq:5}
H=\frac{\kappa_1}3 p^3+\frac{\kappa_0}2 p^2+U_\mathrm{rf}(1
-\cos\phi) \; .
\end{equation}

To analyze a phase portrait of the system, it is expedient to
present Hamilton function of the longitudinal motion in the 
reduced form:
\begin{equation} \label{k:1}
\tilde{H} = \mu\frac{\tilde{p}^3}{3} +\frac{\tilde{p}^2}2 +1
-\cos\phi \;,
\end{equation}
where
\begin{subequations}\label{h:5a}
\begin{align}
\tilde{p} &=\sqrt{\frac{\kappa_0}{U_\mathrm{rf}}}p =
\sqrt{\frac{2\pi h \alpha_0 \gamma_s E_0}{eV}}p \; ;\\
\mu^2 &=\frac{\kappa^2_{1}U_\mathrm{rf}}{\kappa^3_{0}}
=\frac{(\alpha_0+\alpha_1)^2 eV}{2\pi h\alpha^3_0 \gamma_s E_0}\;.
\end{align}
\end{subequations}

\begin{figure*} [bth]
\centering %
\includegraphics[width=\textwidth]{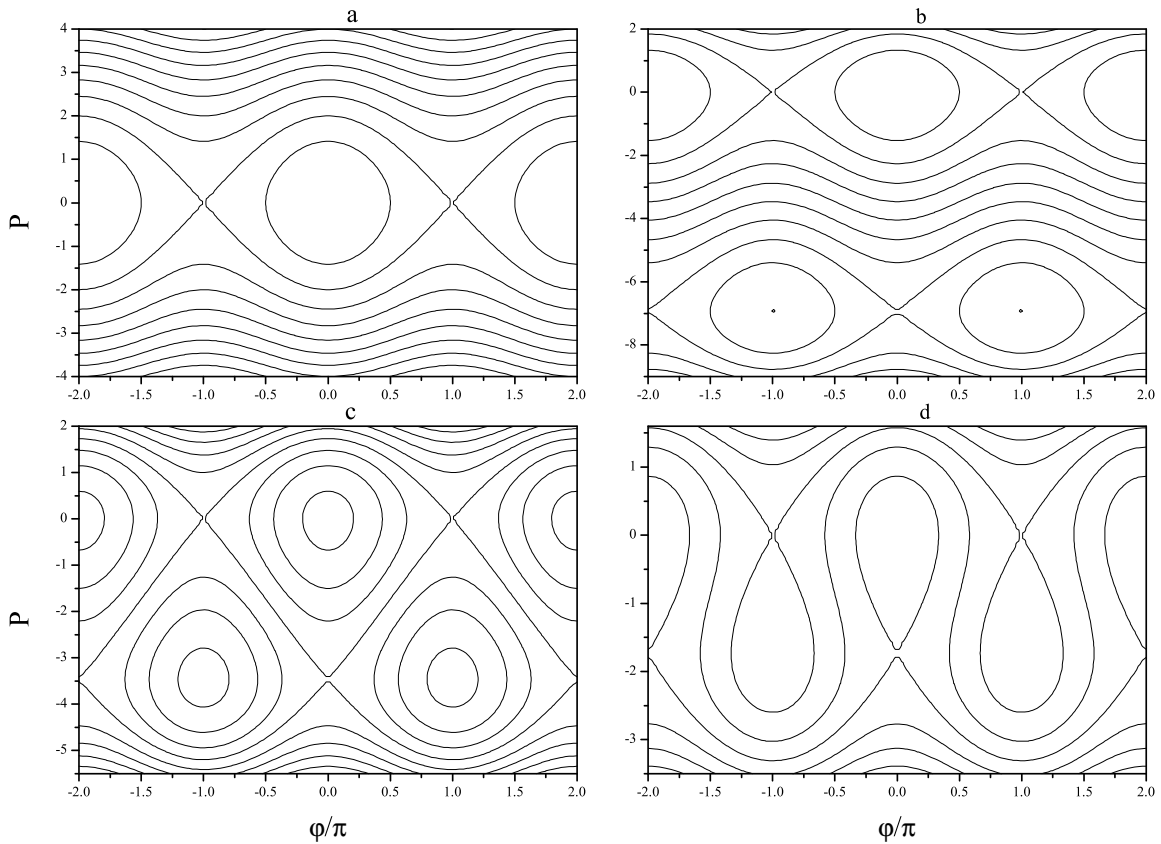}%
\caption[]{Phase portrait of longitudinal motion with account for
the cubic nonlinearity at different values of the parameter $\mu$.
(a): $\mu = 0$, (b): $\mu \leq \left|\mu_{c}\right|$, (c): $\mu =
\mu_{c}$, (d): $\mu \geq \left| \mu_{c} \right| $ \label{fig:1}}
\end{figure*}

Phase portraits of motion with the Hamiltonian \eqref{k:1}
represented in Fig.~\ref{fig:1}. Topology of the phase plane is
governed by the magnitude and sign of the parameter $\mu$. At zero
value, $\mu=0$, the Hamiltonian \eqref{eq:5} or \eqref{k:1} has a
form of mathematical pendulum; its  phase plane is presented in
Fig.~\ref{fig:1}(a).

Within the interval $0\leq \mu^2 < 1/12$, there an additional area
of finite motion appears; this area is separated from the main area
with the band of infinite motion as depicted in Fig.~\ref{fig:1}(b).
When the parameter $\mu $ exceeds the critical value
$\mu^2_{c}=1/12$ [see Fig.~\ref{fig:1}(c)], e.g. $1/12\leq \mu^2 <
\infty $, the structure of the phase plane will have changed as is
represented in Fig.~\ref{fig:1}(d).

The dimension of a stable (finite) longitudinal motion, i.e., the
area comprised by a separatrix, is in direct proportion with ratio
of the ring parameters. For the considered case of the nonlinear
Hamiltonian \eqref{k:1}, the separatrix height (size along the $p$
axis) is determined by
\begin{subequations} \label{k:2}
\begin{align}
\Delta p &= \frac{\alpha_0}{\alpha_0+\alpha_1}
\left[\cos\frac\xi3+\cos\left(\frac\xi3+\frac\pi3\right)\right]
\;,\label{k:2a}\\
&\cos\xi = 12U_\mathrm{rf}\frac{\left(\alpha_0+\alpha_1\right)^2}
{\pi h\alpha^3_0}-1\;, \nonumber\\
\Delta p &= \frac32
\frac{\alpha_0}{\alpha_0+\alpha_1}\;,\label{k:2b}
\end{align}
\end{subequations}
for $\mu\leq\mu_c$ \eqref{k:2a} and $\mu\geq\mu_c$ \eqref{k:2b},
respectively.

The phase width of the separatrix (dimension along the  $\phi$ axis)
is determined by expressions
\begin{subequations}\label{k:3}
\begin{align}
\Delta \phi &= 2\pi\; ; \qquad \mu\le \mu_c\; ,\\
\Delta \phi &= 2\arccos\left[1-\frac{\pi h}{3U_\mathrm{rf}}
\frac{\alpha^3_0}{\left(\alpha_0+\alpha_1\right)^2}\right]\;;
\\
&\mu\geq \mu_c \; ,\nonumber
\end{align}
\end{subequations}
for the subcritical and overcritical values of the parameter $\mu$.

\begin{figure} [bth]
\centering %
\includegraphics[width=\columnwidth]{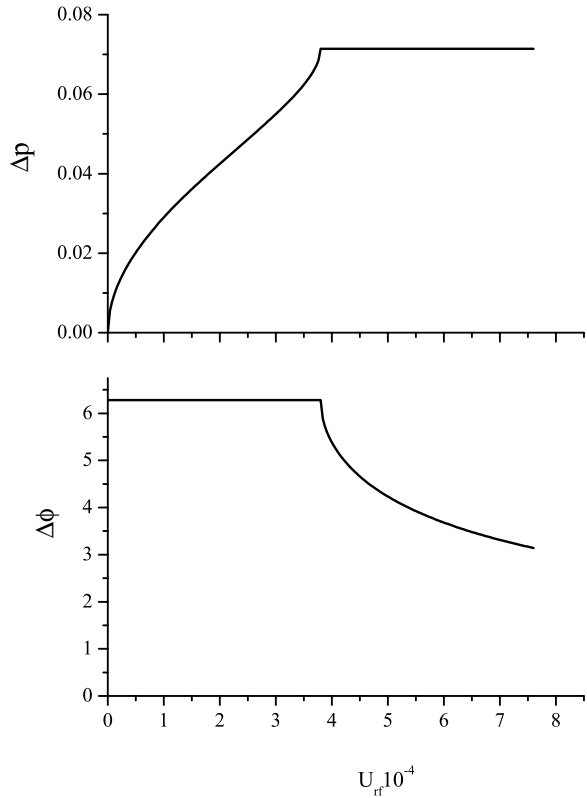}%
\caption[]{Separatrix height (above) and width (below) as functions
of the parameter $U_\mathrm{rf}$ \label{fig:2}}
\end{figure}

Dependence of the phase and momentum separatrix extensions on rf
amplitude at fixed other parameters, which values are listed in
Tab.~\ref{tab:table1}, is presented in Fig.~\ref{fig:2}.

\begin{table} [bth]
\caption{\label{tab:table1} Ring parameters} 
\begin{tabular}{llr}
\hline parameter & desig& value \\ \hline
Accel. voltage (Volt) &$V_\mathrm{rf}$ &  $4\times10^5$\\
Lorentz factor & $\gamma_s$ &  84\\
Harmonic number &$h$ &  32\\
Lin. comp. factor &$\alpha_0$  &  0.01\\
Quad. comp. factor &$\alpha_1$ &  0.2\\
\hline
\end{tabular}
\end{table}

As it can be seen from the plot in Fig.~\ref{fig:2}, while
increasing the parameter $U_\mathrm{rf}$ the separatrix height grows
up reaching its maximum, $\Delta p\approx 7.1 \times 10^{-2}$, at
$U_{\mathrm{rf}(c)}\approx 3.8\times 10^{-4}$ (which is equal to the
rf voltage of $V_{(c)}\approx 16.3$\,kV at $\gamma_s = 84$).

With further increase in the rf voltage, the separatrix height
remains constant.

The separatrix width remains constant with increase of the rf
voltage up to the critical value $U_{\mathrm{rf}(c)}$, then it
\emph{is diminishing.}

\begin{figure} [bth]
\centering %
\includegraphics[width=\columnwidth]{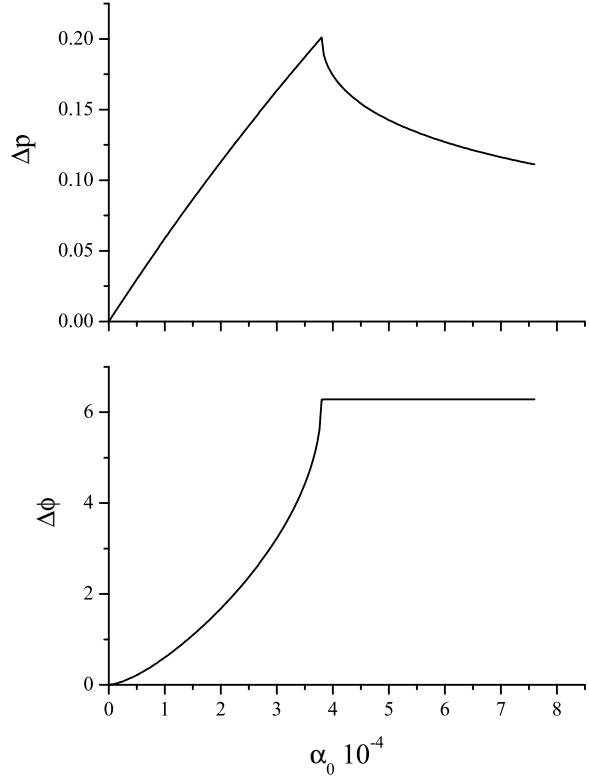}%
\caption[]{Separatrix height (above) and width (below) as functions
of $\alpha_0$ \label{fig:3}}
\end{figure}

In Fig.~\ref{fig:3}, a dependence of the separatrix dimensions upon
the linear momentum compaction factor under other system parameters
fixed is presented.

Quite the reverse to the dependence $\Delta\phi=\Delta\phi
(U_\mathrm{rf})$, a dependence of the separatrix width upon the
linear compaction factor, $\Delta\phi=\Delta\phi(\alpha_{0})$, is
increasing while $\alpha_0$ grows. At a certain critical value of
the linear momentum compaction factor $\alpha_{0(c)}$ (in the
suggested case $\alpha_{0(c)} \approx 0.03$ ), the width of
equilibrium area has reached its maximum and remains constant with
further increase in $\alpha_0$. A dependence of the separatrix
height on $\alpha_0$ is of increasing within interval
$0\leq\alpha_0\leq\alpha_{0(c)}$. Then, after the maximum at
$\alpha_0=\alpha_{0(c)}$ this dependence becomes declining, coming
to zero at a large $\alpha_0$.

Since the phase volume enclosed within the separatrix (and,
therefore, the storage ring acceptance) is proportional to product
of the transverse dimensions of the separatrix, $\sigma\sim \Delta p
\Delta \phi$, then from comparison of the plots in Fig.~\ref{fig:2}
and Fig.~\ref{fig:3} it follows that optimal working point is about
the critical parameters.

In addition, it can be seen that, dislike a linear lattice,
nonlinear terms in the momentum compaction factor restrict the
infinite increase of energy acceptance with decreasing of the linear
momentum compaction factor: The acceptance increase takes place
while the linear compaction is above certain critical value
$\alpha_{0(c)}$, which is determined by the ring lattice parameters
according to equality
\begin{equation} \label{k:4}
\frac{(\alpha_0+\alpha_1)^2 eV}{2\pi h\alpha^3_0 \gamma_s E_0}
=\frac{1}{12}\;.
\end{equation}

With further decrease of $\alpha_{0}$ the acceptance decreases also.

\begin{figure} [bth]
\centering %
\includegraphics[width=\columnwidth]{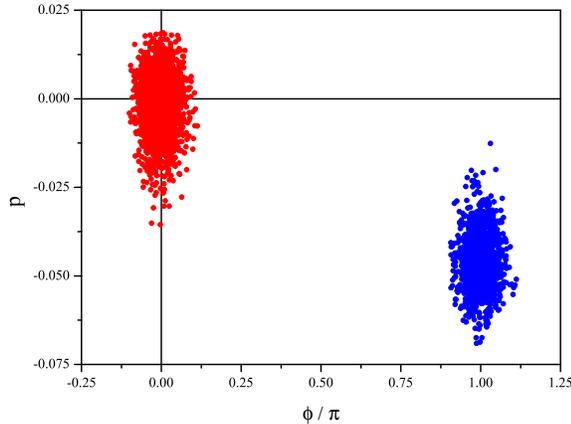}%
\caption[]{Distribution of confined electrons over the longitudinal
phase plane in a system with cubic nonlinearity at $\mu\geq\mu_{c}$;
left bunch corresponds to the ``linear'' case, right -- to the
``nonlinear'' (additional) one \label{fig:4}}
\end{figure}

To validate use of differential (smoothed) equations of motion
\eqref{eq:4} for analysis of Compton storage rings, a code has been
developed based on the finite difference equations \eqref{eq:3}. A
simulated phase space portrait at the ring parameters listed in
Tab.~\ref{tab:table1} for $\mu\geq\mu_{c}$ is presented in
Fig.~\ref{fig:4}.

From the figure it follows that the electrons can be confined within
not only the ``linear'' area (minimum of Hamilton function
\eqref{k:1}), but the ``nonlinear'' as well. (The nonlinear stable
region disappears in a linear lattice.) RMS sizes and the center of
weight positions perfectly correspond to the analytical estimations
presented above.


\section{Summary. Conclusion}
Results of the study on dynamics of synchrotron motion of particles
in the storage rings with the nonlinear momentum compaction factor
presented in the paper, can be digested as follows:
\begin{itemize}
\item Grounded on a simplified model of the storage ring, the
finite-difference equations were derived. Hamiltonian treatment of
the phase space structure was performed. As was shown, the structure
of the phase space is governed by ratios of the ring parameters. An
analytical expression for the factor $\mu $, which determines the
topology of the longitudinal phase space, was derived.

\item Dependencies of the sizes of the equilibrium areas of
the synchrotron  motion in a nonlinear lattice were derived.
Analysis of dependence of the longitudinal acceptance upon the
amplitude of rf voltage, and the linear compaction factor at the
fixed quadratic nonlinear term was presented. As was shown, the
acceptance is growing up only to a definite magnitude which
determines by the critical value of parameter $\mu=\mu_{c}$. It was
emphasized that in order to maximize the acceptance of a lattice
with a small linear momentum compaction factor and a wide energy
spread of electrons in the bunches, the system parameters should be
chosen close to the critical value of $\mu$.

\item To validate the use of smoothed equations of motion, a
simulating code was developed. The code is based on the
finite-difference equations. The results of simulation manifest a
good agreement with the theoretical predictions on the sizes and
position of equilibrium areas.
\end{itemize}

The results obtained allow to make the following conclusion:
Enlargement of the energy acceptance of a ring by decreasing of the
momentum compaction factor is limited with the nonlinearity in the
compaction factor. Decreasing of the linear compaction factor below
the certain limit causes the reversed effect --- decreasing of the
acceptance.

Similar consequence corresponds to the build--up of the rf voltage:
Increase of the voltage above a certain limit causes narrowing of
possible bunch lengthes while the energy acceptance remains
constant. This effect can lead to decrease in the injection
efficiency for high rf voltages.

\end{document}